\documentstyle[preprint,aps]{revtex}

\begin{document}
\begin{center}
{\large \bf An application of supersymmetric quantum mechanics to a planar
physical system}
\end{center}
\centerline{ R. de Lima Rodrigues$^{a,}$\footnote{Permanent address:
Departamento de Ci\^encias Exatas e da Natureza, Universidade Federal
da Para\'\i ba, 58.900-000 Cajazeiras - PB, Brazil,
Email: rafael@cfp.ufpb.br.
Published in
Phys. Lett. {\bf A287}, 45-49, (2001).},
V. B. Bezerra$^b$ and A. N.
Vaidya$^c$}

\begin{center}
$^a$ Centro Brasileiro de Pesquisas F\'\i sicas\\
Rua Dr. Xavier Sigaud, 150\\
22290-180, Rio de Janeiro-RJ, Brazil
\end{center}

\begin{center}
$^b$ Departamento de F\'{\i}sica, Universidade Federal da Para\'{\i}ba\\
58051-970 Jo\~{a}o Pessoa-PB, Brazil\\
\end{center}

\begin{center}
$^c$ Instituto de F\'{\i}sica,
Universidade Federal do Rio de Janeiro \\
21.945-970 Rio de Janeiro-RJ, Brazil
\end{center}

\centerline{\bf Abstract}
\bigskip
Supersymmetry (SUSY)
in non-relativistic quantum mechanics (QM)
is applied to a 2-dimensional physical system:
a neutron in an external magnetic field. The superpotential and the
two-component wave functions of the ground state are found out.


\newpage

The algebraic technique of supersymmetry in quantum
mechanics (SUSY QM) was first introduced by Witten \cite{W}.
The essential idea
of this formulation is based on the Darboux
procedure on second-order differential equations, which has been successfully
utilized to achieve a supersymmetric generalization
of the harmonic-oscillator raising and lowering operators for shape-invariant
potentials \cite{Gen,Fred}.
The SUSY algebra has also been applied to construct
a variety of new one-parameter families of isospectral supersymmetric
partner potentials in quantum field theory \cite{RPV99}.
The shape-invariance conditions in SUSY  have been independently
generalized for systems
described by two-component wave functions \cite{Ta}.
Recently, we have found a two-by-two matrix superpotential associated to
the linear classical stability from the static
solutions for a system of two coupled real scalar fields in
(1+1)-dimensions \cite{RPV98}.

We also presented an integral
representation for the momentum space Green's function for a neutron in
interaction with a static magnetic field of a straight
current carrying wire, which is also described by two-component wave
functions \cite{RV99}. The SUSY QM formalism was also applied to this
planar physical system in the momentum \cite{Voronin} and
coordinate \cite{PRL95} representations.

In this letter, we consider the notation of Ref. \cite{Voronin}.
However, according
to our developments, we can realize the supersymmetric algebra in
coordinate representation, introducing some transformations in the
original system corresponding to a neutron interacting with the magnetic
field of a linear current carrying conductor, so that we are able
to implement a comparasion with both superpotentials for the cases
corresponding to currents located along $x$ and $z$ directions.

Now, let us consider an electrically neutral spin-$\frac 12$ particle of
mass $M=1$ and
magnetic moment $\mu \vec{\sigma}$ (a neutron) interacting with an
infinite straight wire carrying a current $I$ and located along the
$z$-axis. The magnetic field generated by the wire is given by (we use units
with $c=\hbar =1$)

\begin{equation}
\vec{B}=2I\frac{(-y,x,0)}{(x^2+y^2)},
\end{equation}
where $x$ and $y$ are Cartesian coordinates of the plane perpendicular
to the wire.

The Hamiltonian associated with the physical system is given by

\begin{equation}
H=\frac{\vec{p}^{~\!2}}{2M}+\mu
\vec{\sigma}.\vec{B}=
\frac{\vec{p}^{~\!2}}{2}
+2I\mu \frac{(-y\sigma _1+x\sigma _2)}{(x^2+y^2)},
\end{equation}
where $\vec\sigma=(\sigma_1, \sigma_2, \sigma_3)$
are Pauli matrices.
The motion along the $z$-axis is free and will be ignored in what follows and
in this way we get a two-dimensional problem.

Due to  the translational symmetry in the z-direction, the two-component
wave function $\psi(\rho, k)$ can be written as

\begin{eqnarray}
\psi^{(n_\rho)} (\rho, k)&=&\frac 1{\sqrt{4\pi L}}
\left\{\tilde\psi_1^{(n_\rho,m)}(\rho,k)
e^{im\phi }\left(
\begin{array}{c}
1 \\
0
\end{array}
\right) + \tilde\psi_2^{(n_\rho,m)}(\rho, k)
e^{i(m+1)\phi }\left(
\begin{array}{c}
0 \\
1
\end{array}
\right)\right\} e^{i\frac{2\pi }Lkz}\nonumber \\
{}&& \equiv \left(
\begin{array}{c}
\psi_1^{(n_\rho)}(\rho,k) \\
\psi_2^{(n_\rho)}(\rho,k)
\end{array}
\right),
\label{WF}
\end{eqnarray}
where $n_\rho=0, 1, 2, \cdots$ is the radial quantum number;
$k=0, 1, 2, \cdots$; $m=0, \pm 1,
\pm 2, \cdots$;
 $\rho, \phi, z$  are the usual cylindrical coordinates and
the parameter $L$ is the macroscopic length of the conductor.

Therefore, the Schr\"odinger equation splits up into a system of two
coupled second order differential equations as follows

\begin{eqnarray}
\label{SEQ2}
\frac {1}{\rho} \frac{d}{d\rho} \left(\rho\frac{d}{d\rho}
{\tilde \psi_1^{(n_\rho,m)}} \right)-
\frac{m^2}{\rho^2} {\tilde \psi_1^{(n_\rho,m)}+2}{\tilde E}
{\tilde\psi}_1^{(n_\rho,m)}+
\frac{2F}{\rho} {\tilde\psi}_2^{(n_\rho,m)}&=&0,
\nonumber\\
\frac{1}{\rho} \frac{d}{d\rho} \left( \rho \frac{d}{d\rho}
{\tilde \psi_2^{(n_\rho,m)}} \right)-
\frac{(m+1)^2}{\rho^2}{\tilde \psi_2^{(n_\rho,m)}}+2{\tilde E}
{\tilde \psi_2^{(n_\rho,m)}}+\frac{2F}{\rho}{\tilde \psi_1^{(n_\rho,m)}}
&=&0,
\end{eqnarray}
where
\begin{equation}
\label{F}
F=-\frac{\mu_0 \mu I}{2\pi}
\end{equation}
and

\begin{equation}
{\tilde E}=E-\frac{2\pi k^2}{L^2}.
\end{equation}
Note that Eq. (\ref{SEQ2}) is exactly Eq. (2.8) given in
\cite{Voronin}. Now, using the relation

\begin{equation}
\tilde \psi_i^{(n_\rho,m)}=\rho^{-\frac{1}{2}} \phi_i^{(n_\rho,m)}
\quad  (i=1,2),
\end{equation}
we can write the system in (\ref{SEQ2}) in the matrix form as

\begin{equation}
\left(\begin{array}{ll}
-\frac{d^2}{d \rho^2}+\frac{m^2-\frac{1}{4}}{\rho^2}-
2{\tilde E}  & \;\;\;\;\;\;\frac{-2F}{\rho}\\
\frac{-2F}{\rho} & -\frac{d^2}{d \rho^2}+
\frac{(m+1)^2-\frac{1}{4}}{\rho^2}-2{\tilde E}
\end{array}
\right)
\left(\begin{array}{l}
\phi_1^{(n_\rho,m)}\\
\phi_2^{(n_\rho,m)}
\end{array}
\right)=0,
\end{equation}
which corresponds to a one-dimensional Schr\"odinger-like equation
associated with the two-component wave function. Therefore, we get the
eigenvalue equations

\begin{equation}
{\bf H}_1{\bf\Phi}^{(n_\rho,m)}_1={\tilde E}^{(n_\rho,m)}_1
{\bf\Phi}^{(n_\rho,m)}_1, \quad E^{(n_\rho,m)}_1=2{\tilde E}^{(n_\rho,m)},
\end{equation}
where

\begin{equation}
\label{wfn}
{\bf\Phi}^{(n_\rho,m)}_1={\bf\Phi}^{(n_\rho,m)}_1(\rho,k)=\left(
\begin{array}{c}
\phi_1^{(n_\rho,m)}(\rho,k) \\
\phi_2^{(n_\rho,m)}(\rho,k)
\end{array}
\right)
\end{equation}

and

\begin{equation}
\label{wfn1}
{\bf H_1}=-{\bf 1}\frac{d^2}{d \rho^2}+
\left(\begin{array}{llll}
\frac{m^2-\frac{1}{4}}{\rho^2} & \;\;\;\frac{-2F}{\rho}\\
\frac{-2F}{\rho} &\frac{(m+1)^2-\frac{1}{4}}{\rho^2}
\end{array}
\right).
\end{equation}
Defining

\begin{equation}
\label{sf}
{\bf H}_1\equiv {\bf A}^{+}{\bf A^{-}}+
{\bf 1}{\tilde E}^{(0)}_1,
\quad {\bf A}^{\pm}=\pm\frac{d}{d\rho}+{\bf W}(\rho),
\end{equation}
we obtain the following Riccati equation in matrix form

\begin{equation}
\mbox{\bf W}^{\prime}(\rho)+\mbox{\bf W}^2(\rho)+
\mbox{\bf 1}{\tilde E}^{(0)}_1
=\left(\begin{array}{llll}
\frac{m^2-\frac{1}{4}}{\rho^2} & \;\;\;\frac{-2F}{\rho}\\
\frac{-2F}{\rho} &  \frac{(m+1)^2-\frac{1}{4}}{\rho^2}
\end{array}
\right),
\end{equation}
where  $\mbox{\bf W}(\rho )$ is a two-by-two superpotential matrix.
The hermiticity condition allows us to write

\begin{equation}
\mbox{\bf W}=\mbox{\bf W}^{\dagger}
=\left(\begin{array}{ll}
f(\rho)  &  g(\rho)\\
g(\rho)  &  h(\rho)
\end{array}
\right),
\end{equation}
where $f, g$ and $h$ are real functions and satisfy the
 nonlinear system of differential equations

\begin{eqnarray}
\label{snl3}
\left\{\begin{array}{lll}
f^{\prime}+f^2+g^2+E^{(0)}_1&=&
\frac{m^2-\frac{1}{4}}{\rho^2}\\
fg+hg+g^{\prime}&=&\frac{-2F}{\rho}\\
h^{\prime}+h^2+g^2+E^{(0)}_1&=&
\frac{(m+1)^2-\frac{1}{4}}{\rho^2}.
\end{array}\right.
\end{eqnarray}

Now, let us try a solution for equation (\ref{snl3}) assuming that $g$ is
constant. Then, we have

\begin{equation}
f+h=-\frac{2F}{g\rho},
\end{equation}
which gives

\begin{equation}
\label{e16}
f^{\prime}-h^{\prime} -\frac{2F}{g\rho}\left(f-h\right)+
\frac{2m+1}{\rho^2}=0.
\end{equation}
Solving the last equation and imposing finiteness condition on the
solutions, we get

\begin{eqnarray}
\label{e17}
f(\rho)=&&\frac{b}{\rho}, \nonumber\\
h(\rho)=&&\frac{c}{\rho},
\end{eqnarray}
where $b$ and $c$ are arbitrary constants.
Substituting these solutions into the system (\ref{snl3}), we find
that a consistent solution is possible only if

\begin{equation}
\label{e18}
g=-\frac{F}{m+1}
\end{equation}
where $F$ is defined in Eq. (\ref{F}). Then, turning to Eq. (\ref{e16})
and substituting Eqs. (\ref{e17}) and (\ref{e18}) we find constants $b$ and
$c$. Putting these results back into Eq. (\ref{e17}), we have that
\begin{eqnarray}
\label{E16}
f(\rho)&=&\frac{m+\frac{1}{2}}{\rho}\nonumber\\
h(\rho)&=&\frac{m+\frac{3}{2}}{\rho}.
\end{eqnarray}
In this case, the two
almost isospectral  Hamiltonians are given by

\begin{eqnarray}
\label{hnf1}
\hbox{{\bf H}}_1&=&
\hbox{{\bf A}}^+ \hbox{{\bf A}}^--\frac{F^2}{2(m+1)^2}\hbox{{\bf 1}},\\
\hbox{{\bf H}}_2
{}&&= \hbox{{\bf A}}^- \hbox{{\bf A}}^+ -\frac{F^2}{2(m+1)^2}\hbox{{\bf 1}}.
\end{eqnarray}
Since $\hbox{{\bf A}}^+ \hbox{{\bf A}}^-$  is positive
semidefinite, according to (\ref{sf}) and (\ref{hnf1}) the energy
eigenvalue of the ground state is

\begin{equation}
{\tilde E}^{(0)}=-\frac{F^2}{2(m+1)^2},
\end{equation}
with the annihilation conditions

\begin{equation}
\label{AC1}
A^- \Phi^{(0)}_1 = 0
\end{equation}
and

\begin{equation}
\label{AC2}
A^+ \Phi^{(0)}_2 = 0
\end{equation}
and the new superpotential

\begin{equation}
\mbox{\bf W}_m=\left(\begin{array}{ll}
\frac{m+\frac{1}{2}}{\rho}  &  -\frac{F}{m+1}\\
-\frac{F}{m+1}  & \frac{m+\frac{3}{2}}{\rho}
\end{array}
\right).
\end{equation}

The energy eigenvalues of magnetically bound excitated states in
terms of the radial quantum number $n_\rho$, for $m\geq m_0$ becomes

\begin{equation}
{\tilde E}^{(n_\rho)}=-\frac{F^2}{2(n_\rho + m_0 +1)^2}.
\end{equation}

Now let us to determine the eigenfunction associated with the ground state
given by Eq.({\ref{AC1}). To do this let us consider the transformations

\begin{equation}
\phi(\rho)=\chi^{(0)}\rho^{m+\frac 12}, \quad \rho = 2(m+1)\eta,
\quad F=-\frac 12,
\end{equation}
which implies that Eq. (\ref{AC1}) turns into the following matrix diferential
equation

\begin{equation}
\label{hn}
{\bf 1}\frac{d}{d \eta}\chi(\eta)=
\left(\begin{array}{llll}
0 & 1\\
1 &  \frac{1}{\eta}
\end{array}
\right)\chi^{(0)}(\eta ), \quad
\chi^{(0)}(\eta)=\left(
\begin{array}{l}
\chi^{(0)}_1\\
\chi^{(0)}_2
\end{array}
\right),
\end{equation}
so that we obtain the following equations for
the components $\chi^{(0)}_1$ and $\chi^{(0)}_2:$

\begin{eqnarray}
\frac{d}{d\eta}\chi^{(0)}_1(\eta)&&=
\chi^{(0)}_2(\eta), \nonumber\\
\frac{d}{d\eta}\chi^{(0)}_2(\eta)&&=
\chi^{(0)}_1(\eta)+\frac{1}{\eta}\chi^{(0)}_2(\eta),
\label{X2}
\end{eqnarray}
which leads us a second-order differential equation for
$\chi^{(0)}_2(\eta),$ viz.,

\begin{equation}
\label{DEX2}
\frac{d^2}{d\eta^2}\chi^{(0)}_2(\eta)-\frac{1}{\eta}
\frac{d}{d\eta}\chi^{(0)}_2(\eta)+
\left(\frac{1}{\eta^2}-1\right)\chi^{(0)}_2(\eta)=0.
\end{equation}

From equations (\ref{WF}), (\ref{X2}) and (\ref{DEX2}) we obtain the
$m$-dependent normalizable ground state

\begin{equation}
\Psi^{(0)}(\rho)=
C_m\rho^{m+1}\left(\begin{array}{l}
e^{im\phi}K_1\left(\frac{\rho}{2m+2}\right)\\
e^{i(m+1)\phi}K_0\left(\frac{\rho}{2m+2}\right)
\end{array}
\right)e^{i\frac{2\pi}{L}kz}
\end{equation}
where $C_m$ is the normalization constant,
and $K_1\left(\frac{\rho}{2m+2}\right)$ and
$K_0\left(\frac{\rho}{2m+2}\right)$ are the modified Bessel functions.
The eigenfunction $\Psi^{(0)}(\rho)$ is in according with the result
found via momentum representation in Ref. \cite{Voronin}.
Note that the complete solution of Eq. (\ref{DEX2})

\begin{equation}
\chi^{(0)}_2(\eta)= \eta\left(C_1K_0(\eta)+C_2I_0(\eta)\right),
\end{equation}
where $C_1$ and $C_2$ are arbitrary non-normalizable constants .
Therefore, in order to get a normalizable solution, we choose $c_2=0$
and in this way we drop $I_0(\eta)$ which is divergent when
$\eta\rightarrow \infty.$

It is worthy noticing that under a unitary transformation,
$
{\bf U}{\bf W}_m{\bf U}^{-1}={\tilde{\bf W}}_m,
$
this  superpotential, together with the interchange of $m$ by $m+\frac 12,$
and taking $F=-\frac 12$ becomes that  superpotential
matrix $({\bf W}_{LJM})$ shown in \cite{PRL95},
viz., ${\tilde{\bf W}}_{m+\frac 12}=-{\bf W}_{LJM}.$
This minus sign that connects ${\tilde{\bf W}}_{m+\frac 12}$
and ${\bf W}_{LJM}$ is associated to the fact that we have
chosen the first-order differential operator ${\bf A}^-$
with the opposite sign in the derivative term of the operator $A_m$
considered in Ref. \cite{PRL95}.

Using the coordinate representation,
we investigate the SUSY in non-relativistic quantum mechanics with
two-component eigenfunctions and find a new realization of
supersymmetry in a planar physical system of a neutron in
interaction with a straight current-carrying wire.

The $N=2-$SUSY superalgebra has  the following
representation

\begin{equation}
\label{E40}
{\bf H}_{SUSY} = [Q_- ,Q_+ ]_+ = \left(
\begin{array}{cc}
\mbox{\bf A}^+ \mbox{\bf A}^- & 0 \\
0 & \mbox{\bf A}^-\mbox{\bf A}^+
\end{array}
\right)_{4\hbox{x}4}= \left(
\begin{array}{cc}
{\bf H}_-={\bf H} & 0 \\
0 & {\bf H}_+
\end{array}
\right),
\end{equation}
where the supersymmetric partners are given by
${\bf H}_-={\bf H}_1-\mbox{\bf 1}E^{(0)}_1, \quad
{\bf H}_+={\bf H}_2-\mbox{\bf 1}E^{(0)}_1$
and the supercharges $Q_{\pm}$ are 4 by 4 matrix differential operators
of first order and can be given by

\begin{equation}
\label{sc2}
Q_- = \left(
\begin{array}{cc}
0 & 0 \\
\mbox{\bf A}^- & 0
\end{array}
\right)_{4\hbox{x}4}, \quad
Q_+ = \left(
\begin{array}{cc}
0 & \mbox{\bf A}^+\\
0 & 0
\end{array}
\right)_{4\hbox{x}4}.
\end{equation}

We have seen that, in non-relativistic quantum mechanics
applied to two-component eigenfunctions,
if $\Phi^{(0)}_1$ is a
normalizable two-component eigenfunction,
one cannot write $\Phi^{(0)}_2$ in terms of $\Phi^{(0)}_1,$ as in the
case of ordinary supersymmetric quantum mechanics. This
may be shown in the system considered here of a neutron interacting with  an
external magnetic field with the current of the conductor in the $z$
direction. Only in the case of 1-component wave functions one may
write the superpotential as
$W(x)=\frac{d}{dx}\ell n \left(\psi_0(x)\right)$.

The hermiticity condition satisfied by the
superpotential, in the general case, leads us to a method that permits
to solve the matrix Riccati equation. As a final remark, we would like to
draw the attention to the fact that our result, for a superpotential
corresponding
to a neutron in an external magnetic field in the coordinate representation,
is related by the following unitary transformation,
$
{\bf U}=\frac{1}{\sqrt{2}}\left(\sigma_1+\sigma_3\right),
$
where $\sigma_1$ and $\sigma_3$ are the Pauli matrix,
with a new superpotential so that,
after the substituition $m$ by $m+\frac 12$ (the total angular
momentum along the wire direction) it reduces to the
superpotential recently found in \cite{PRL95}, where a current
$I$ along the $x$ axis of a Cartesian system is considered.

\noindent
{\bf Acknowledgments}

\noindent
This research was supported in part by CNPq (Brazilian Research Agency).
RLR wish to thanks the staff of the CBPF and DCEN-CFP-UFPB for the facilities.
Thanks are also due to J. A. Helayel Neto for hospitality of RLR
at CBPF-MCT and for fruitful discussions on supersymmetric models.

\end{document}